# ADAPTIVE OPTICS FOR MULTIFOCAL PLANE MICROSCOPY

Benjamin Gore[1,2*], Noah Schwartz[1], Paul Dalgarno[2]

[1]STFC UK Astronomy Technology Centre, Royal Observatory, Edinburgh EH9 3HJ, United Kingdom [2] Institute of Biological Chemistry, Biophysics and Bioengineering, *Heriot-Watt University, Edinburgh, EH14 4AS, United Kingdom* [*]Corresponding Author: Benjamin.gore@stfc.ac.uk

## ABSTRACT

Multifocal plane microscopy (MUM) allows three dimensional objects to be imaged in a single camera frame. Our approach uses dual orthogonal diffraction phase gratings with a quadratic distortion of the lines to apply defocus to the first diffraction orders which, when paired with a relay lens, allows for 9 focal planes to be imaged on a single camera chip. This approach requires a strong signal level to ensure sufficient intensity in the diffracted light, but has the advantage of being compact and straightforward to implement. As the microscope begins to focus deeper into the sample, aberrations caused by refractive index mismatch and inhomogeneity in the sample's media have an adverse effect on the signal's quality.

In this paper, we investigate the image quality improvement brought by applying adaptive optics (AO) to multifocal plane microscopy. A single correction device (an 8x8 deformable mirror (DM)) is combined with an image-based AO control strategy to perform the correction of optical aberrations. We compare full end-to-end modelling results using an established numerical modelling system adapted for microscopy to laboratory results both on a test sample and on a number of biological samples. Finally, we will demonstrate that combining AO and MUM, we are able to improve the image quality of biological samples and provide a good correction throughout the volume of the biological sample.

## 1. INTRODUCTION

The challenge of wavefront sensing in microscopy requires a different approach to astronomy where many of the first adaptive optics techniques were developed. The rapidly changing nature of atmospheric turbulence means that measurement and correction of the aberrations should be accomplished at a high repetition rate. Due to the distances involved, objects are generally considered to be at infinity, meaning all objects in the image are in focus, and for a localized region (i.e. smaller than the anisoplanatic patch) in the sky the aberrations can be considered to be uniform [1]. In the case of microscopy its a three-dimensional sample being viewed. This results in light from out of focus parts of the sample encroaching on the image, making wavefront sensing using traditional methods (e.g. Shack-Hartmann using simple centre of gravity algorithms) problematic [2]. This is exacerbated by the fact that the aberrations experienced are also depth dependent. However, the approximately static nature of the aberrations means that new options for wavefront sensing become viable. Optimization-based approaches are generally preferred for microscopy by means of maximizing a chosen image-based metric. This optimization can be challenging as the large number of degrees of freedom mean that conventional approaches can take a long time to converge. Over-exposure of a fluorescent sample leads to photo-bleaching, so it is desirable to reduce the convergence time of the optimization process as much as possible.

This paper will aim to apply adaptive optics to volumetric imaging techniques, specifically multifocal plane microscopy through the use of a modified diffraction grating [3]. Multifocal techniques require the light to be split into multiple paths, which reduces the SNR of the final images. Correcting for aberrations, particularly those caused by imaging deep into the sample, will improve the viability of multiplane imaging and hence increase the number of potential applications.

## 2. MULTIFOCAL PLANE MICROSCOPY

For a diffraction grating with parallel lines, light will experience a change in direction dependant on the diffraction order. By introducing a quadratic distortion into the lines, a positive defocus is introduced into the positive diffraction orders, and a negative defocus in the negative orders, as in an off-axis Fresnel lens. By pairing the grating with a lens in a 4f system, multiple image planes can be captured simultaneously with a single camera. The result is a compact subsystem which can simply be placed in front of the microscope camera.

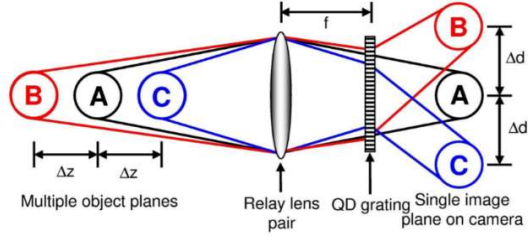

**Figure 1: Schematic of a quadratically distorted (QD) diffraction grating sub-system. Image planes separated by an amount Δz and shown on the camera separated by an amount Δd [4]**

## 3. NUMERICAL SIMULATION RESULTS

Simulation work has been carried out using the Object-Orientated Matlab Adaptive Optics (OOMAO) toolbox [5], which has been modified for use in microscopy by creating additional objects. This has allowed the viability of AO correction techniques to be investigated, acts as a baseline performance estimation for future experimental work, and allows examination of the effect the multiplane system has on image correction. Simulating the multiplane system was achieved via simulating a number of beads at various depths in focus simultaneously. The spherical aberration due to depth was added to each bead [8], and the aberration induced by the sample was based on the Kolmogorov (following an f-11/3 power spectrum) model as previously recommended in Schmitt et. Al. [9]. The deeper beads encountered more layers of Kolmogorov turbulence than the shallower ones.

The primary technique being investigated is the image sharpening technique developed by Booth [6], an image-based optimization technique which has been found to significantly improve convergence times on traditional algorithms. Applying a known aberration mode to an image and recording the response of a suitable metric function shows an approximately Lorentzian function. After extensive analysis, a spectral density metric based on the optical transfer function was selected [7]. By taking a series of images with varying amounts of the chosen aberration mode applied, a suitable function can be fitted to the data in order to obtain the peak, which will occur when the mode is corrected. By applying this technique to each mode sequentially the wavefront can be estimated in a modal fashion.

The number of samples taken per mode and the spacing between these samples were investigated as variables. It was found that 3 samples per mode was viable but unstable. 5 samples provided reliable performance, and more yielding diminishing returns. The optimal spacing was found to be impossible to discern as it was heavily dependent on the exact metric response curve and the magnitude of the aberrations. For noisy and high aberration systems larger spacing was preferred.

Optimizing the signal using the image sharpening method could be carried out on either an entire or partial image. This allowed the algorithm to optimize for a certain depth and then view the effects of applying the correction on the other layers, as well as viewing the effects of optimizing the entire image. The spectral density metric for each bead was recorded for each of these scenarios as each mode was corrected sequentially.

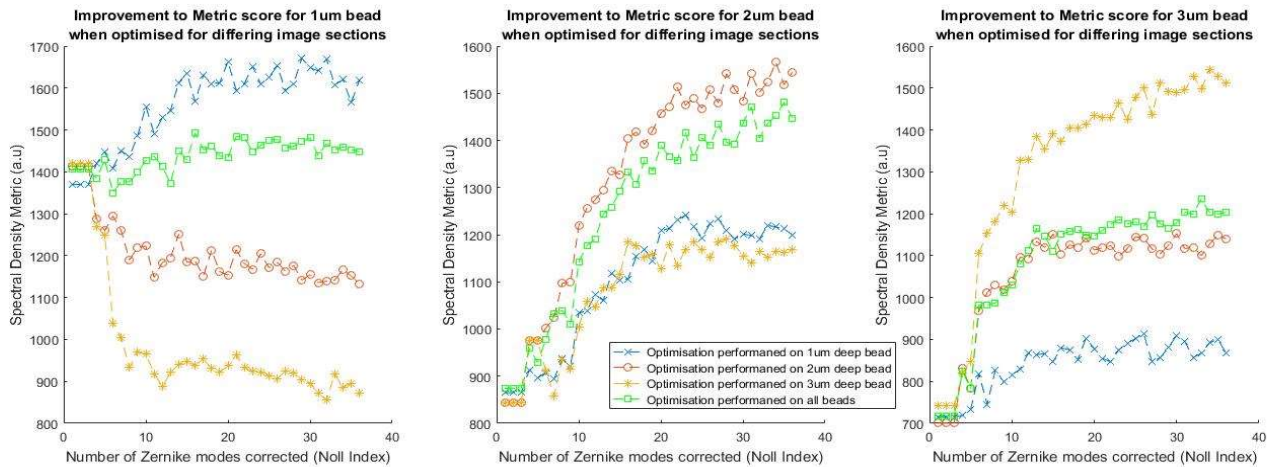

**Figure 2: Three beads at various depths were simulated with a spherical aberration proportional to its depth and accumulative layers of Kolmogorov turbulence above each bead. The metric score for each bead was monitored as the optimisation was carried out. Shown are the impact to the metric score of the three beads for the 4 corrected strategies**

Basing the optimization on the entire image can induce a quality of correction that can be comparable to the correction achieved via optimizing each layer independently (fig 2). Extending this analysis to 9 planes yields similar results; all planes are improved with the exception of the top-most plane experiencing a slight drop in signal level (fig 3). These results confirm the viability of implementing a single correction device. Although further improvements could be found with individual devices for each layer to fine tune the performance this would also introduce many practical difficulties.

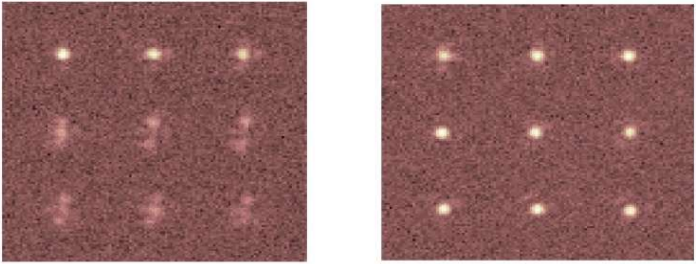

Figure 3: (Left) 9 in focus beads simulated at different depths with depth-dependant aberrations and accumulative Kolmogorov turbulence layers. (Right) The 9 beads after optimising based on the entire image. All beads bar the top-most plane experience an increase in metric score.

## 4. EXPERIMANTAL WORK

The deformable mirror was positioned within an optical relay attached to the side port of the microscope (Olympus IX71S8F) camera port. Its reflection was passed through the diffraction grating sub-assembly before being imaged onto the camera. The diffraction grating can easily be slid in and out of the system, allowing a quick transition between regular imaging and MUM. Due to thermal effects and membrane memory effects in the DM [5], achieving repeatable surface performance was found to be problematic for open loop operation. As a result, a secondary closed loop optical path allows the shape of the mirror to be continually monitored using a Shack-Hartmann wavefront sensor to ensure the mirror is able to converge to the desired shape, reducing the residual error between repeat DM commands by up to 30x. The influence function of the DM was measured using a Zygo interferometer, and an analysis through OOMAO found the DM was unable to accurately produce many higher order Zernike terms.

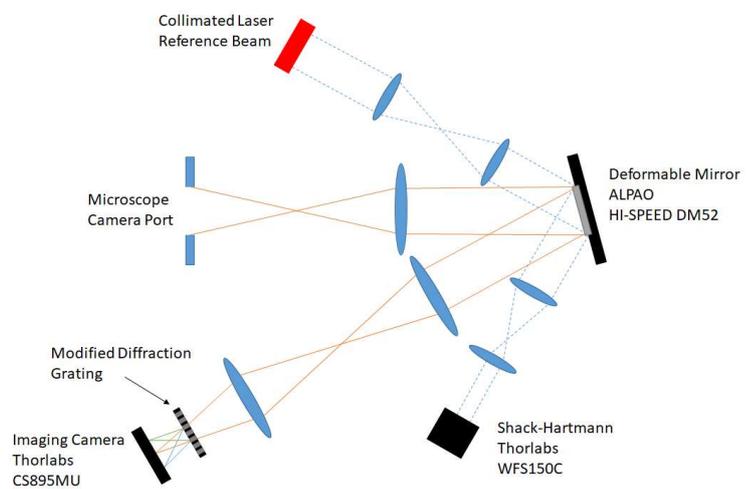

Figure 4: The light from the microscope camera port is transformed onto the DM surface, and its reflection is passed through the modified diffraction grating sub-assembly.

The algorithm was therefore confined to optimising the first 28 modes in order to get the most correction before photo bleaching occurs.

Preliminary results have shown that, even in open loop operation, the system can detect and correct for large induced aberrations. These images were taken by placing pieces of transparent plastic in the imaging path whilst imaging florescent bead samples. We use fluorescent bead samples with aberrations introduced into the imaging path to simulate additional sample aberrations. Preliminary results using the experimental setup have demonstrated that the system can measure and correct for large aberrations introduced by the systems and the samples. Despite the major challenge of operating the DM in open loop, a significant image quality improvement can be seen (figure 5, left). Quantitatively, the image metric was increased by over 50% after optimisation.

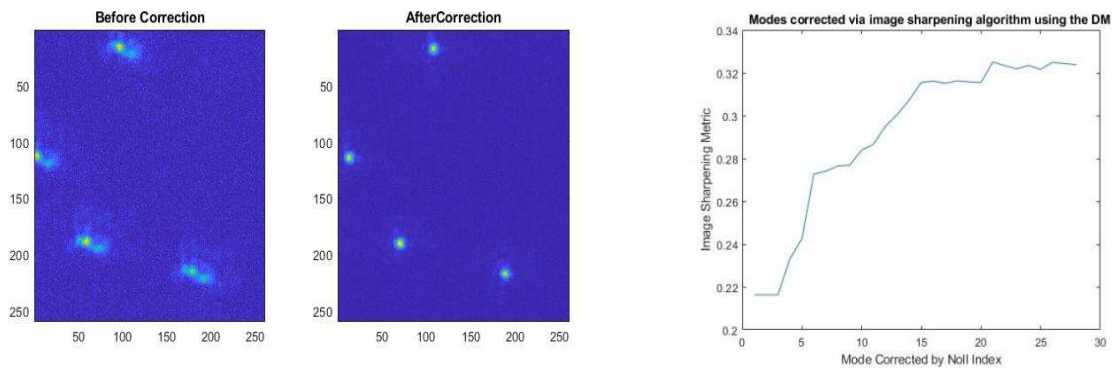

Figure 5: (Left) Fluorescent bead samples with aberrations introduced into the imaging path. (Right) Metric score per iteration of the correction algorithm.

Initial tests were performed on biological tissues samples. We were only able to obtain a slight increase in image metric value, primarily due to minimal aberrations form our thin test samples. However this proves the methodology is robust and enables a pathway to exploring more aberration effected 3D cell cultures and neurological samples as benchmark systems.

## 5. CONCLUSIONS

In this paper, we presentation simulation and experimental results for adaptive optics correction on a multifocal plan microscope. We first validated the correction approach using a mature end-to-end numerical simulation tool; and finally presented some preliminary laboratory experimental results. We have shown a very good match between the simulated results and the experimental data. One of the major technical issues we had to overcome was the open loop performance of the deformable mirror. A secondary closed loop system was introduced to mitigate these effects, with the SH-WFS loop causing the residual error between repeat DM commands to be reduced by up to 30x. Open loop correction was performed on bead samples with a visible improvement of image quality and a metric improvement by more than 50%. Work is currently underway to increase the close-loop (i.e. SH-WFS and DM) speed and procure suitable biological tissue samples to be corrected using the DM in conjunction with the multiplane imaging system.

## REFERENCES

[1] R. K. Tyson, Principles of adaptive optics, 2nd Edition, San Diego: Academic Press, 1998.

[2] M. J. Booth, "Adaptive optical microscopy: the ongoing quest for a perfect image," *Light: Science & Applicationsvolume 3,* vol. e165, no. 3, 2014.

[3] P. Blanchard and A. Greenaway, "Simultaneous multiplane imaging with a distorted diffraction grating," *Applied Optics,* vol. 38, no. 32, pp. 6692-6699, 1999.

[4] P. Dalgarno, H. Dalgarno, A. Putoud, R. Lambert, L. Paterson, D. Logan, D. Towers, R. Warburton and A. Greenaway, "Multiplane imaging and three dimensional nanoscale particle tracking in biological microscopy," *OPTICS EXPRESS,* vol. 18, no. 2, p. 887, 2010.

[5] U. Bitenc, N. A. Bharmal, M. T. J. and R. M. Myers, "Assessing the stability of an ALPAO deformable mirror for feed-forward operation," *Optics Express,* vol. 22, no. 10, pp. 12438-12451, 2014.

[6] M. J. Booth, "Wavefront senorless adaptive optics for large aberrations," *OPTICS LETTERS,* vol. 32, no. 1, pp. 5-7, 2007.

[7] J. M. Schmitt and G. Kumar, "Turbulent nature of refractive-index variations in biological tissue," *Optics Letters,* vol. 21, no. 16, pp. 1310-1312, 1996.

[8] P. Kner, J. W. Sedat, D. A. Agard and Z. Kam, "High-resolution wide-field microscopy with adaptive optics for spherical aberration correction and motionaless focusing," *Journal of Microscopy,* vol. 237, no. 2, pp. 136-147, 2009.

[9] R. Conan and C. C, "Object-oriented Matlab adaptive optics toolbox," in *Proceedings Volume 9148, Adaptive Optics Systems IV;*, Montréal, 2014.

[10] D. Debarre, M. J. Booth and T. Wilson, "Imagebased adaptive optics through optimisation of low spatial frequencies," *Optics Express,* vol. 15, no. 13, pp. 8176-8190, 2007.